\documentclass{svproc}
\usepackage{url}
\usepackage{cite}
\usepackage{amsmath,amssymb,amsfonts}
\usepackage{algorithmic}
\usepackage{graphicx}
\usepackage{textcomp}
\usepackage{xcolor}
\usepackage{graphicx}
\usepackage{cite}
\usepackage{amsmath,amssymb,amsfonts}
\usepackage{algorithmic}
\usepackage{graphicx}
\usepackage{textcomp}
\usepackage{xcolor}
\usepackage{comment}
\usepackage{hyphenat}
\usepackage{multirow}
\usepackage{tabu}
\usepackage{caption}
\usepackage{subcaption}
\usepackage{
tikz,
relsize,
amsmath,
booktabs,
tikz
}
\usetikzlibrary{arrows,calc,positioning,shadows,shapes}
\usepackage{pgfplots}
\usepackage{pgfplotstable}
\pgfplotsset{compat=1.6}

\usepackage[T1]{fontenc}
\usepackage[utf8]{inputenc}
\usepackage{lmodern}

\usepackage[
labelfont=sf,
hypcap=false,
format=hang,
width=0.8\columnwidth
]{caption}

\usetikzlibrary{
calc,trees,shadows,positioning,arrows,chains,
decorations.pathreplacing,
decorations.pathmorphing,
decorations.shapes,
decorations.text,
shapes,
shapes.geometric,
shapes.symbols,
matrix,
patterns,
intersections,
fit}

\pgfdeclarelayer{background layer}
\pgfdeclarelayer{foreground layer}
\pgfsetlayers{background layer,main,foreground layer}

\tikzset{
>=latex
}

\usepackage[utf8]{inputenc}
\usepackage{amsmath}
\usepackage{amssymb}
\usepackage{arydshln}
\usepackage{mathrsfs}
\usepackage[export]{adjustbox}
\usepackage{float}
\usepackage{times}
\usepackage{fancyhdr,graphicx,amsmath,amssymb}
\usepackage[ruled,vlined]{algorithm2e}

\begin{document}
\mainmatter              
%
\title{The Ties that matter: From the perspective of Similarity Measure in Online Social Networks}


%
\titlerunning{The Ties that matter: From the perspective of Similarity Measure in Online Social Networks}  
%
\author{Soumita Das \and Anupam Biswas
}
\authorrunning{Das and Biswas} 
%
\tocauthor{Soumita Das, Anupam Biswas}
\institute{Department of Computer Science and Engineering,\\
National Institute of Technology, Silchar-788010, Assam, India  \\
\email{wingsoffire72@gmail.com},\\ 
\email{anupam@cse.nits.ac.in}}

\maketitle              

\begin{abstract}
Online Social Networks have embarked on the importance of connection strength measures which has a broad array of applications such as, analyzing diffusion behaviors, community detection, link predictions, recommender systems. Though there are some existing connection strength measures,  the density that a connection shares with it's neighbors and the directionality aspect has not received much attention. In this paper, we have proposed an asymmetric edge similarity measure namely, Neighborhood Density-based Edge Similarity (NDES) which provides a fundamental support to derive the strength of connection. The time complexity of NDES is $O(nk^2)$. An application of NDES for community detection in social network is shown. We have considered a similarity based community detection technique and substituted its similarity measure with NDES. The performance of NDES is evaluated on several small real-world datasets in terms of the effectiveness in detecting communities and compared with  three widely used similarity measures. Empirical results show NDES enables detecting comparatively better communities both in terms of accuracy and quality.

\end{abstract}

\section{Introduction}
\label{one}

Significant advancements in the usage of Online Social Networks (OSNs) have flooded the internet with unprecedented volume of connections. The users of OSNs make connections based on common interests and backgrounds. With the rapid growth of connections in the cyberspace, the OSNs have become dynamic in nature. In order to emphasize the evolution of these networks, it is necessary to predict the probability of a social connection between two users. This is widely referred as the link prediction problem ~\cite{liben2007link, palla2005uncovering, lu2011link, al2011survey, martinez2016survey,dunlavy2011temporal}. Links can be predicted through similarity between two users based on various features. Recently, local similarity measures have gained a lot of attention because of it's accuracy and low time complexity. Some of the local similarity measures include Common neighbors (CN)~\cite{liben2007link},  Jaccard index~\cite{jaccard1901etude}, Adamic Adar (AA)~\cite{adamic2003friends},  Resource allocation (RA)~\cite{zhou2009predicting}, Sorensen~\cite{sorensen1948method}, Salton Index (SA)~\cite{salton1983introduction}, Mutual Information (MI)~\cite{tan2014link}, preferential attachment~\cite{barabasi2002evolution}, measure by H. Ghorbanzadeh et al.~\cite{ghorbanzadeh2021hybrid}  and so on. Local similarity measures based on node degree and common neighbors information have been explored by V. Mart{\'\i}nez et al.~\cite{martinez2016survey}.  However, conventional local similarity measures such as common neighbors, adamic adar, resource allocation, resource allocation based on common neighbor interactions fails to normalize it's similarity score. In fact, such measures could not make a distinction between edge scores for directed edges. Aghabozorgi et al.~\cite{aghabozorgi2018new} proposed a local measure based on a combination of common neighbors and triadic structures blocks to obtain similarity between two users but their approach is not practical for large networks due to high time complexity. The local measure presented by Li et al. ~\cite{li2018similarity} is based on future common neighbors, where neighbors are classified into three classes.

Finding similarity between connected nodes can be used to detect group of similar nodes commonly referred as a community. Particularly, community detection algorithms based on local similarity have low time complexity~\cite{jiang2013efficient, li2015parameter, wang2015novel}. Intuitively, these algorithms are mostly seed expansion oriented where at first seed nodes are selected based on certain criterion and then communities are expanded by adopting similarity measure. For instance, $(\alpha, \beta)$ algorithm explores the role of Jaccard Index in the detection of communities~\cite{eustace2015community}, another community detection approach incorporates random walk to traverse neighborhood information matrix~\cite{zhou2003distance, clauset2005finding }. Moreover, local similarity can also be deployed in label propagation algorithm~\cite{bagrow2005local, raghavan2007near}. Next, it is very important to validate the performance of the local similarity measures. To do so, we have selected a similarity based community detection method namely, Closeness Similarity driven Information Diffusion based community
detection (CSID)~\cite{das2021community}. We incorporated our proposed local similarity measure i.e. Neighborhood  Density-based Edge Similarity (NDES) to generate communities. Next, the communities are evaluated in two perspectivies: quality and accuracy. Quality metrics are used to determine the structural feasibility of detected communities by considering connectivity within and outside a community. Whereas, accuracy metrics are deployed to evaluate the accuracy of the detected communities by comparing with ground-truth~\cite{das2021deployment}. Different quality and accuracy metrics are available to evaluate the goodness and correctness of detected communities respectively.  We have evaluated NDES measure on CSID algorithm by considering several small real-world datasets, popular local simiarity measures, accuracy metrics and quality metrics. Comparative analysis of the results obtained would assist to infer the quality and accuracy of our proposed measure. 

In this paper, a novel asymmetric edge similarity measure has been presented. The proposed measure considers neighborhood density to compute edge similarity. The contributions of this paper are as follows:

\begin{itemize}
    \item Introduced a novel neighborhood density based edge similarity measure considering neighborhood information. To do so, concepts such as neighborhood density, information, common neighbors and node degree has been utilized. 
    
    \item Importance of information associated with each node for computation of edge similarity score have been addressed. This information possessed by a node has been deployed to normalize our similarity score. This normalization part is important to reduce redundancy and improve score accuracy.

    \item Edge similarity score obtained by implementation of our proposed measure is asymmetric because of the incorporation of normalization. Therefore, it is possible to compute edge similarity score for directed graphs as well.

    \item Evaluation of communities obtained by incorporation of our proposed similarity measure on a certain community detection algorithm suggest that proposed similarity measure gives comparatively good results both in terms of accuracy and quality. 
    
    \item Works equally well in both dense and sparse networks and the time complexity is relatively less as compared to the representative local similarity measures.
    
\end{itemize}

The rest of the paper is organized as follows: Section 2
defines the proposed local similarity measure,
Section 3 gives a literature review about existing local similarity measures, Section 4 presents a detailed discussion about our proposed approach, Section 5 discusses about the experimental analysis part and Section 6 is about conclusion.


\section{Problem Definition}
\label{two}
Given a simple graph $G (V, E)$ where $V$ indicates set of nodes, ~$E$ indicates set of edges. Each edge refers to a connection between node pairs $\{x,y\}$  where $x,y \in V$. The problem of edge similarity measure  $s_{x,y}$ is to find proximity between connected node pairs based on local connectivity pattern. It is generally influenced by attributes such as neighbors, common neighbors and degree of the nodes under consideration because nodes share strong connections only with those nodes with which it shares tight neighborhood.  The greater the value contributed by these attributes, higher is the $s_{x,y}$ score.


Often, similarity scores of all edges are computed to predict link establishment. Mostly, local similarity based problems consider common neighbors to determine $s_{x,y}$ scores and are symmetric in nature. Moreover, most of the existing measures are not normalized. It is to be mentioned here that normalization is important to reduce redundancy and improve accuracy. We introduce the notion of neighborhood density,  information (see Section ~\ref{four}) in addition to common neighbors.  Neighborhood density plays a significant role to derive the similarity between connected nodes because highly similar entities are densely connected with their neighboring entities. Additionally, it is not compulsory for both the nodes sharing a connection to have the same affinity towards each other. Therefore, we have introduced the information measure to normalize our proposed method. Thus, for a given graph $G$, the new local similarity measure utilizes common neighbor, neighborhood density and information to compute the similarity score.

\section{Related Work}
\label{three}

In general, link prediction approaches are categorized into three classes: similarity based algorithms, maximum-likelihood based algorithms, probabilistic models~\cite{lu2011link}. Similarity based algorithms are based on computation of the similarity score $s_{xy}$ between node pairs $(x,y)$. Higher $s_{x,y}$ value indicates higher probability of link establishment. In maximum-likelihood based algorithms, the link establishment is calculated based on a series of rules. Whereas, in probabilistic models, an abstract model of the network is generated to determine link establishment. Out of these algorithms, similarity based algorithms are comparatively simple and have low time complexity and therefore, applicable on large scale networks as well. Recently, local similarity based measures received a lot of attention due to it's low time complexity, effectiveness and relative accuracy in social networks. 

To explore the local similarity measures, let us first formalize some of the mostly used terms. Suppose, we have a graph $G(V, E)$ comprising of $n$ number of nodes and $m$ number of edges. For a node say $x \in V$,  set of neighbors and degree of node $x$ is represented by $\Gamma(x)$ and $\mid \Gamma_{x} \mid$ respectively. If we assume $\{x,y\}$ to be a connected node pair where $x,y \in V$, then set of common neighbors and number of common neighbors between nodes $x$ and $y$ is indicated by $\Gamma(x) \cap \Gamma(y)$ and $\mid \Gamma(x) \cap \Gamma(y) \mid$ respectively. Union of node $x$ and node $y$ indicated by $\Gamma(x) \cup \Gamma(y)$ and $k$ indicates largest node degree of graph $G$. Some of the existing local similarity measures are summarized as follows:\\

\textbf{Adamic Adar:} This measure has been primarily proposed to calculate the similarity between two web pages and is defined by~\cite{adamic2003friends},
\begin{equation}
    \label{aa}
    s_{xy} = \sum_{z \in \Gamma(x)\cap \Gamma(y)}\frac{1}{log(\Gamma(z)}
\end{equation}

From equation~\ref{aa}, it is inferred that this is not a normalized measure and is symmetric. It's time complexity is $O(nk^{3})$.\\

\textbf{Preferential Attachment:} It is defined by~\cite{barabasi2002evolution},

\begin{equation}
    \label{pa}
    s_{xy} = \mid \Gamma_{x}\mid \times \mid \Gamma_{y} \mid
\end{equation}

From equation~\ref{pa}, it is inferred that it is not a normalized similarity score and is symmetric. It's time complexity is $O(nk^{3})$.\\

\textbf{Resource Allocation:} Both adamic adar and resource allocation measures penalizes common neighbors having high degree. However, RA gives better efficiency for networks with high average degrees. RA is defined as~\cite{zhou2009predicting},

\begin{equation}
    \label{ra}
    s_{xy} = \sum_{z\in \Gamma(x) \cap \Gamma(y)}\frac{1}{\Gamma(z)}
\end{equation}

From equation~\ref{ra}, it can be derived that it not not a normalized measure and is symmetric. It's time complexity is $O(nk^{3})$.\\

\textbf{Salton Index:} It is also referred as cosine measure and is defined as~\cite{salton1983introduction},

\begin{equation}
\label{si}
    s_{xy} = \frac{\mid \Gamma(x) \cap \Gamma(y) \mid}{\sqrt{\mid \Gamma_{x} \mid \times \mid \Gamma_{y}\mid} }
\end{equation}

Equation~\ref{si} indicates that it is a normalized similarity score and is symmetric.  It's time complexity is $O(nk^{3})$.\\

\textbf{Jaccard Coefficient:} It is defined by~\cite{jaccard1901etude},
\begin{equation}
\label{jjss}
    s_{xy} = \frac{\mid \Gamma(x) \cap \Gamma(y) \mid}{\mid \Gamma(x) \cup \Gamma(y)\mid }
\end{equation}
Equation~\ref{jjss} indicates that it is a normalized similarity score and is symmetric.  It's time complexity is $O(nk^{3})$.

\textbf{Discussion:} In this paper, we have developed a novel edge similarity measure called NDES to estimate the proximity of vertices in a network. Existing similarity measures do not address the directionality and time complexity aspect properly. NDES has been developed to address the above mentioned issues. The NDES approach utilizes neighborhood density score to analyze similarity shared by connected node pairs. Here,  neighborhood density is used to identify density of connections shared by connected node pairs. This is beneficial in this context because larger number of connections between connected node pair indicates higher similarity between nodes under consideration.



\section{Proposed Approach}
\label{four}


In this section, we shall discuss about our proposed local edge similarity measure which is based on density of connections shared by an edge with it's neighbors. To do so, at first we formalize some basic concepts such as neighborhood density,  number of common neighbors, degree of node that are used to design our proposed edge similarity measure. Additionally, we have designed a new information quantification formula to compute information possessed by a node for normalizing our edge similarity measure. As the information attached with connected nodes may be different, so the edge similarity score given by our proposed method is asymmetric.

\subsection{Preliminaries}
\begin{itemize}

\item{Neighborhood  density:} For a connected node pair $\{x,y\}$, neighborhood density is used to compute the density of it's neighbors. This neighborhood density score is utilized to measure the connection similarity of the node pair under consideration. Let us assume that we have connected node pair $\{x,y\}$ where  $x,y  \in V$, then neighborhood density is indicated by $\rho_{x,y}$. Higher $\rho_{x,y}$ value indicates greater connection similarity score.\\


\item{Number of common neighbors:} OSN users have an inclination to bond more with other users who are known to them. These acquaintances are identified by the connectivity information which gives the number of common neighbors shared by an edge. Here, the number of common neighbors shared by an edge (say, $e_{x,y}$) is indicated by $\mid \Gamma(x) \cap \Gamma(y) \mid$. It is used to compute the similarity score associated with edge $e_{x,y}$. The greater the number of common neighbors shared by an edge, larger is the similarity score of the corresponding edge.\\

\item{Degree of node:} It reflects the number of connections associated with a node. It plays a vital role to determine the relationship between two nodes connected by an edge.  For instance, if the degree of a node (say, $x$) indicated by $\mid \Gamma_{x} \mid$ is one, then it is evident that node $x$ shares maximum similarity with it's associated edge.\\

\item {Information $I_{x}$:} It indicates the amount of information related to node $x$ in an undirected and unweighted network $G$. If connected nodes possess equal amount of information, it indicates that they are surrounded by the same set of entities and hence, are similar.

\end{itemize}

\subsection{Neighborhood Density-based Edge Similarity}
We have designed an edge strength measure namely, Neighborhood Density-based Edge Similarity (NDES) considering neighborhood density of the edge for better comprehension of the similarity shared by connected nodes. Suppose, we have connected node pair $\{x,y\}$, where $x, y \in V$ , then we use the term $e_{x,y}$ to represent the edge shared by nodes $x$ and $y$. The similarity score of node pair $\{x,y\}$ is based on common neighborhood computation. In this context, common neighborhood computation is defined in terms of neighbors between nodes $x$ and $y$. Higher common neighborhood score indicates greater interaction frequency and hence, higher similarity. We need to consider following cases for common neighborhood computation.\\


\textbf{Case a.} Nodes $x$ and $y$ do not have any common neighbor i.e. $|\Gamma(x) \cap  \Gamma(y)| = \phi$, then common neighborhood score of connected node pair $\{x,y\}$  is defined by,



\begin{equation}
\label{esz}
\rho_{x,y}  =\left\{\begin{matrix}
1, & if~ \mid \Gamma_{x} \mid ~=1 ~or~\mid \Gamma_{y} \mid=1 \\ 
0, & if ~\mid \Gamma_{x} \mid>1~or~\mid \Gamma_{y} \mid>1
\end{matrix}\right.
\end{equation}

\textbf{Case b.} Nodes $x$ and $y$ shares one or more than one common neighbor i.e. if $|\Gamma(x) \cap  \Gamma(y)| >= 1$, then common neighborhood score of connected node pair $\{x,y\}$ is defined by,





\begin{equation}
\label{eso}
\resizebox{.994\hsize}{!}{$\rho_{x,y}  =  |\Gamma(x) \cap \Gamma(y)| + |\Gamma(x) \cap \Gamma(z)|  +  |\Gamma(y) \cap \Gamma(z)| +  |\sigma_{xy}| + |\Gamma(w)  \cap \Gamma(z)|,~~  \forall w,z \in (\Gamma(x) \cap \Gamma(y)) ~if~ e_{w,z} \in E, ~w \ne z. $} 
\end{equation}

Equation~\ref{eso} consists of five parts. Here, the first term $\mid \Gamma(x) \cap \Gamma(y) \mid$ indicates common neighbors shared by nodes $x$ and $y$, symbols $\mid \Gamma(x) \cap \Gamma(z) \mid$ and  $\mid \Gamma(y) \cap \Gamma(z) \mid$ in the second and third term indicates number of common neighbors shared by nodes $x$ and $y$ with common neighbor of $x$ and $y$ respectively, forth term $\mid \sigma_{xy}  \mid$ indicates number of connections shared by common neighbors of $x$ and $y$,  fifth term ~~$\mid \Gamma(w)  \cap \Gamma(z) \mid$ indicates number of common neighbors shared by common neighbors of $x$ and $y$. Next, after the common neighborhood score $\rho_{x,y}$ have been computed for all the edges, we try to address the directionality aspect associated with an edge by computing the information possessed by a node say $x$ as $I_{x}$ which is defined as,

\begin{equation}
    \label{info}
    I_{x} = max_{y \in \Gamma(x)}\rho_{x,y}
\end{equation}
Equation~\ref{info} indicates that information possessed by node $x$ is equal to the maximum neighborhood density score that $x$ shares with it's neighboring nodes. Next, NDES score of connected node pair $\{x,y\}$ is computed by,

\begin{equation}
    \label{NDES}
    NDES (x,y) = \frac{\rho_{x,y}}{I_{x}}
\end{equation}

And NDES score of an edge $(y,x)$ is computed by the following equation,

\begin{equation}
    \label{NDESr}
    NDES (y,x) = \frac{\rho_{x,y}}{I_{y}}
\end{equation}

If $\rho_{x,y}= 0$, then NDES $(x,y)=0$.

\subsection{Time Complexity:}
It can be inferred from NDES measure description that there are three parts associated with this measure. In the first part,  finding common neighbors takes $O(n)$ time complexity. Next, in the second part, time complexity of finding possible connections among common neighbors and number of common neighbors shared by connected common neighbors is around $O(nk^{2})$ where $k$ refers to maximum degree of the network. In the third part, the time complexity associated with finding information associated with all the nodes is $O(n)$. Therefore, total time complexity associated with NDES is $O(nk^{2})$. It is to be mentioned here that our method need not be applied in complete graphs because it won't be meaningful to find similarity scores of node pairs in completely connected graphs. Therefore, total time complexity is  $O(nk^{2})$.

\section{Experimental Analysis}
\label{five}
\pgfplotstableread[row sep=\\,col sep=&]{
Edge similarity measures & NDES & Cosine & Jaccard & Adamic adar\\
NMI & 0.904 & 0.520 & 0.000000000000000436 & 0.000000000000000436 \\
ARI & 0.850 & 0.190 & 0.0 & 0.0\\
NF1 & 0.601 & 0.167 & 0.21 & 0.21 \\
}\lookm

\begin{figure}
\centering
\begin{subfigure}{0.51\textwidth}
\centering
\begin{tikzpicture}[remember picture]
    \begin{axis}[
            ybar,
           bar width=.22cm,
            width=\textwidth,
            height=\textwidth,
            enlarge x limits=0.2,
            legend style={at={(0.5,1)},
                anchor=north,legend columns=2,legend cell align=left},
            symbolic x coords={NMI,ARI,NF1},
            xtick=data,
             x tick label style={rotate=00,anchor=north},
            nodes near coords align={vertical},
            ymin=-0.10,ymax=2.0,
            ylabel={Edge similarity measures},
            nodes near coords,
            every node near coord/.append style={rotate=90, anchor=west}
        ]
        \addplot table[x=Edge similarity measures,y=NDES]{\lookm};
        \addplot table[x=Edge similarity measures,y=Cosine]{\lookm};
        \addplot table[x=Edge similarity measures,y=Jaccard]{\lookm};
        \addplot table[x=Edge similarity measures,y=Adamic adar]{\lookm};
         \legend{NDES,Cosine,Jaccard,Adamic adar}
    \end{axis}
\end{tikzpicture}
\captionsetup{skip=10pt}
\caption{Strike}
\end{subfigure}%
\pgfplotstableread[row sep=\\,col sep=&]{
Edge similarity measures & NDES & Cosine & Jaccard & Adamic adar\\
NMI & 1.0 & 0.329 & 0.000000000000000642 & 0.000000000000000642 \\
ARI & 1.0 & 0.139 & 0.0 & 0.0\\
NF1 & 1.0 & 0.076 & 0.345 & 0.345 \\
}\lookm
\begin{subfigure}{0.51\textwidth}
\centering
\begin{tikzpicture}[remember picture]
    \begin{axis}[
            ybar,
           bar width=.22cm,
            width=\textwidth,
            height=\textwidth,
            enlarge x limits=0.2,
            legend style={at={(0.5,1)},
                anchor=north,legend columns=2,legend cell align=left},
            symbolic x coords={NMI,ARI,NF1},
            xtick=data,
             x tick label style={rotate=00,anchor=north},
            nodes near coords align={vertical},
            ymin=-0.10,ymax=2.0,
            ylabel={Edge similarity measures},
            nodes near coords,
            every node near coord/.append style={rotate=90, anchor=west}
        ]
        \addplot table[x=Edge similarity measures,y=NDES]{\lookm};
        \addplot table[x=Edge similarity measures,y=Cosine]{\lookm};
        \addplot table[x=Edge similarity measures,y=Jaccard]{\lookm};
        \addplot table[x=Edge similarity measures,y=Adamic adar]{\lookm};
         \legend{NDES,Cosine,Jaccard,Adamic adar}
    \end{axis}
\end{tikzpicture}
\captionsetup{skip=10pt}
\caption{Karate}
\end{subfigure}
\caption{Comparative accuracy analysis of edge similarity measures on Strike and Karate dataset}
\label{figone}
\end{figure}
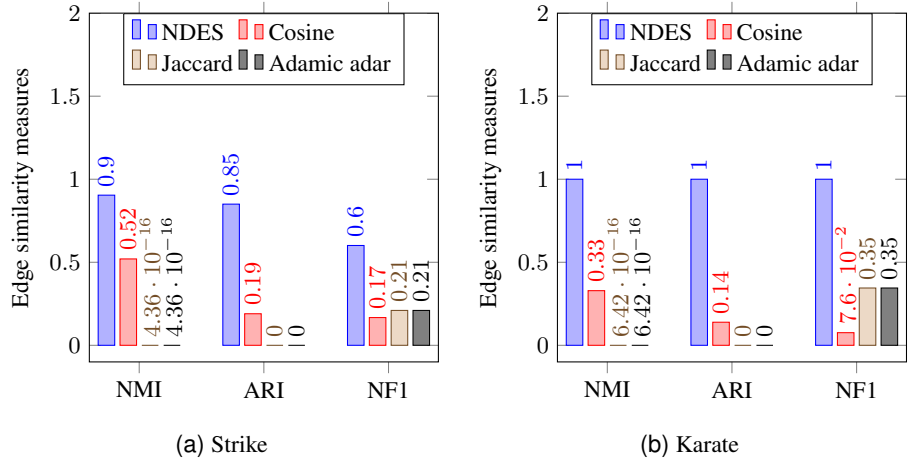


\pgfplotstableread[row sep=\\,col sep=&]{
Edge similarity measures & NDES & Cosine & Jaccard & Adamic adar\\
NMI & 0.834 & 0.503 & 0.452 & 0.0 \\
ARI & 0.626 & 0.184 & 0.272 & 0.0 \\
NF1 & 0.232 & 0.158 & 0.183 & 0.223 \\
}\lookm
\begin{figure}
\centering
\begin{subfigure}{0.51\textwidth}
\centering
\begin{tikzpicture}[remember picture]
    \begin{axis}[
            ybar,
           bar width=.22cm,
            width=\textwidth,
            height=\textwidth,
            enlarge x limits=0.2,
            legend style={at={(0.5,1)},
                anchor=north,legend columns=2,legend cell align=left},
            symbolic x coords={NMI,ARI,NF1},
            xtick=data,
             x tick label style={rotate=00,anchor=north},
            nodes near coords align={vertical},
            ymin=-0.10,ymax=2.4,
            ylabel={Edge similarity measures},
            nodes near coords,
            every node near coord/.append style={rotate=90, anchor=west}
        ]
        \addplot table[x=Edge similarity measures,y=NDES]{\lookm};
        \addplot table[x=Edge similarity measures,y=Cosine]{\lookm};
        \addplot table[x=Edge similarity measures,y=Jaccard]{\lookm};
        \addplot table[x=Edge similarity measures,y=Adamic adar]{\lookm};
         \legend{NDES,Cosine,Jaccard,Adamic adar}
    \end{axis}
\end{tikzpicture}
\captionsetup{skip=10pt}
\caption{Riskmap}
\end{subfigure}%
\pgfplotstableread[row sep=\\,col sep=&]{
Edge similarity measures & NDES & Cosine & Jaccard & Adamic adar\\
NMI & 0.722 & 0.468 & 0.0 & 0.0 \\
ARI & 0.635 & 0.119 & 0.0 & 0.0\\
NF1 & 0.088 & 0.079 & 0.135 & 0.135 \\
}\lookm
\begin{subfigure}{0.51\textwidth}
\centering
\begin{tikzpicture}[remember picture]
    \begin{axis}[
            ybar,
           bar width=.22cm,
            width=\textwidth,
            height=\textwidth,
            enlarge x limits=0.2,
            legend style={at={(0.5,1)},
                anchor=north,legend columns=2,legend cell align=left},
            symbolic x coords={NMI,ARI,NF1},
            xtick=data,
             x tick label style={rotate=00,anchor=north},
            nodes near coords align={vertical},
            ymin=-0.10,ymax=2.4,
            ylabel={Edge similarity measures},
            nodes near coords,
            every node near coord/.append style={rotate=90, anchor=west}
        ]
        \addplot table[x=Edge similarity measures,y=NDES]{\lookm};
        \addplot table[x=Edge similarity measures,y=Cosine]{\lookm};
        \addplot table[x=Edge similarity measures,y=Jaccard]{\lookm};
        \addplot table[x=Edge similarity measures,y=Adamic adar]{\lookm};
         \legend{NDES,Cosine,Jaccard,Adamic adar}
    \end{axis}
\end{tikzpicture}
\captionsetup{skip=10pt}
\caption{Dolphin}
\end{subfigure}
\caption{Comparative accuracy analysis of edge similarity measures on Riskmap and Dolphin dataset}
\label{figtwo}
\end{figure}
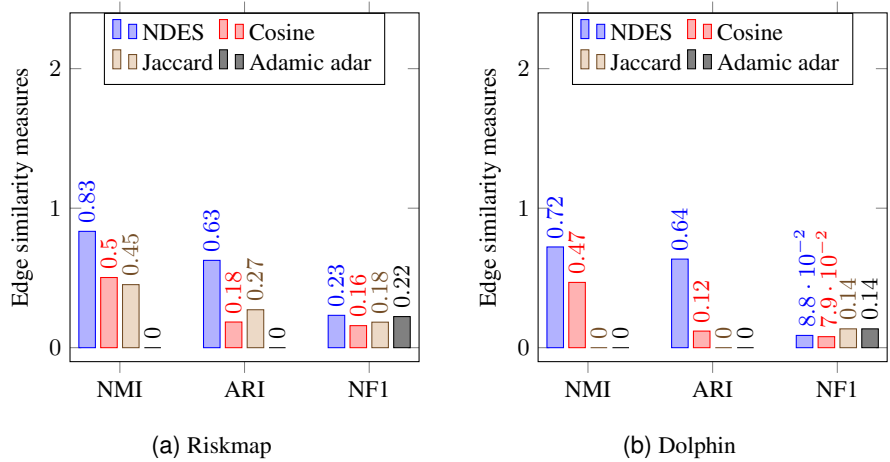

In this section, the strategy for evaluating the performance of NDES is detailed. We have selected a community detection algorithm namely, Closeness Similarity driven Information Diffusion based community detection (CSID) and substituted their similarity measure namely, Closeness Similarity Measure~\cite{das2021community} with popular local similarity  measures such as, Salton Index~\cite{salton1983introduction},  Jaccard Index~\cite{jaccard1901etude}, Adamic Adar~\cite{adamic2003friends} and our proposed similarity measure namely, NDES one by one as shown in Table~\ref{similarity} to perform the comparative analysis. The communities obtained by incorporating these measures are evaluated in two directions:  accuracy and quality to infer about the efficiency and effectiveness of our proposed similarity measure. Accuracy metrics are used to evaluate the correctness of detected communities using Normalized Mutual Information (NMI), Adjusted Random Index (ARI) and Normalized F1-Score (NF1). Furthermore, quality metrics are used to evaluate structural feasibility of detected communities which in turn is used to comparatively evaluate the goodness of our selected and proposed similarity measures. We have selected quality metrics based on  internal connections such as modularity, external connections such as cut\_ratio, expansion and combination of internal and external connections such as conductance~\cite{chakraborty2017}. Next, several datasets  are selected which are publicly available SNAP~\cite{snap} repository. The quality and accuracy results obtained by incorporation of each of the similarity measures on CSID on several small real-world datasets such as strike network, karate network, riskmap network and dolphin network having ground-truth are presented in this section.

\begin{table}
\caption {Comparative summarization of local similarity measures}
\label{similarity}
\centering
  \begin{tabular} {| m{10em}| m{8em} | m{6em} |m{6em}|}
   \hline
   \centering{Similarity} & Normalized/ Not normalized & Symmetric/ Asymmetric &  Time Complexity  \\
   \hline
     \centering{NDES}  &\centering{Normalized} & \centering{Asymmetric} & \centering{$O(nk^{2})$} 
   \tabularnewline
     \centering{Salton Index}  &\centering{Normalized} & \centering{Symmetric} & \centering{$O(nk^{3})$} 
   \tabularnewline
   \centering{Jaccard coefficient}  &\centering{Normalized} & \centering{Symmetric} & \centering{$O(nk^{3})$}  
   \tabularnewline
   \centering{Adamic Adar}  &\centering{Not normalized} & \centering{Symmetric} & \centering{$O(nk^{3})$}  
   \tabularnewline
    \hline
\end{tabular}
\end{table}


\pgfplotstableread[row sep=\\,col sep=&]{
Edge similarity measures & NDES & Cosine & Jaccard & Adamic adar\\
Modularity & 0.534 & 0.351 & 0.0000000000000000018 & 0.0000000000000000018 \\
Conductance & 0.204 & 0.536 & 0.0 & 0.0\\
Cut\_ratio & 0.027 & 0.083 & 0.0 & 0.0\\
Expansion & 0.506 & 1.78 & 0.0 & 0.0\\ 
}\lookm

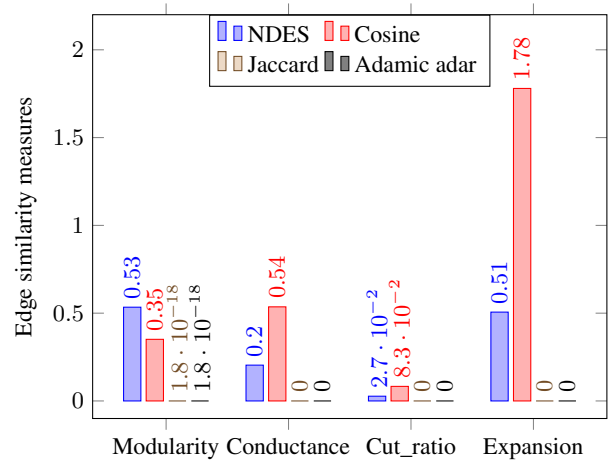
\begin{figure}[!ht]
\centering
\begin{tikzpicture}
    \begin{axis}[
            ybar,
           bar width=.23cm,
            width=0.69\textwidth,
            height=0.57\textwidth,
            enlarge x limits=0.2,
            legend style={at={(0.5,1)},
                anchor=north,legend columns=2,legend cell align=left},
            symbolic x coords={Modularity,Conductance,Cut\_ratio,Expansion},
            xtick=data,
             x tick label style={rotate=00,anchor=north},
            nodes near coords align={vertical},
            ymin=-0.10,ymax=2.2,
            ylabel={Edge similarity measures},
            nodes near coords,
            every node near coord/.append style={rotate=90, anchor=west}
        ]
        \addplot table[x=Edge similarity measures,y=NDES]{\lookm};
        \addplot table[x=Edge similarity measures,y=Cosine]{\lookm};
        \addplot table[x=Edge similarity measures,y=Jaccard]{\lookm};
        \addplot table[x=Edge similarity measures,y=Adamic adar]{\lookm};
         \legend{NDES,Cosine,Jaccard,Adamic adar}
  ,  \end{axis}
\end{tikzpicture}
\caption{\footnotesize Comparative quality analysis of edge similarity measures on Strike dataset}
\label{figthree}
\end{figure}

\pgfplotstableread[row sep=\\,col sep=&]{
Edge similarity measures & NDES & Cosine & Jaccard & Adamic adar\\
Modularity & 0.371 & 0.085 & 0.0 & 0.0 \\
Conductance & 0.128 & 0.0 & 0.0 & 0.0\\
Cut\_ratio & 0.034 & 0.0 & 0.0 & 0.0\\
Expansion & 0.590 & 0.0 & 0.0 & 0.0\\
}\lookm

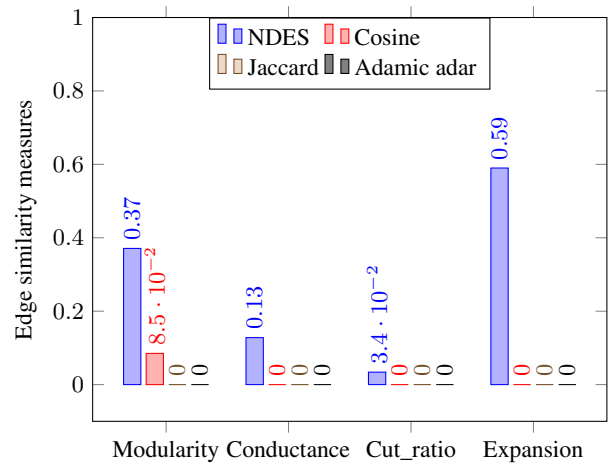
\begin{figure}[!hb]
\centering
\begin{tikzpicture}
    \begin{axis}[
            ybar,
           bar width=.23cm,
            width=0.69\textwidth,
            height=.57\textwidth,
            enlarge x limits=0.2,
            legend style={at={(0.5,1)},
                anchor=north,legend columns=2,legend cell align=left},
            symbolic x coords={Modularity,Conductance,Cut\_ratio,Expansion},
            xtick=data,
             x tick label style={rotate=00,anchor=north},
            nodes near coords align={vertical},
             ymin=-0.10,ymax=1.0,
            ylabel={Edge similarity measures},
            nodes near coords,
            every node near coord/.append style={rotate=90, anchor=west}
        ]
        \addplot table[x=Edge similarity measures,y=NDES]{\lookm};
        \addplot table[x=Edge similarity measures,y=Cosine]{\lookm};
        \addplot table[x=Edge similarity measures,y=Jaccard]{\lookm};
        \addplot table[x=Edge similarity measures,y=Adamic adar]{\lookm};
         \legend{NDES,Cosine,Jaccard,Adamic adar}
  ,  \end{axis}
\end{tikzpicture}
\caption{\footnotesize Comparative quality analysis of edge similarity measures on Karate dataset}
\label{figfour}
\end{figure}

\pgfplotstableread[row sep=\\,col sep=&]{
Edge similarity measures & NDES & Cosine & Jaccard & Adamic adar\\
Modularity & 0.623 & 0.268 & 0.258 & 0.0 \\
Conductance & 0.161 & 0.660 & 0.0 & 0.0\\
Cut\_ratio & 0.017 & 0.06 & 0.0 & 0.0\\
Expansion & 0.629 & 2.32 & 0.0 & 0.0\\
}\lookm

\begin{figure}[!ht]
\centering
\begin{tikzpicture}
    \begin{axis}[
            ybar,
           bar width=.23cm,
            width=0.69\textwidth,
            height=.57\textwidth,
            enlarge x limits=0.2,
            legend style={at={(0.5,1)},
                anchor=north,legend columns=2,legend cell align=left},
            symbolic x coords={Modularity,Conductance,Cut\_ratio,Expansion},
            xtick=data,
             x tick label style={rotate=00,anchor=north},
            nodes near coords align={vertical},
             ymin=-0.10,ymax=3.0,
            ylabel={Edge similarity measures},
            nodes near coords,
            every node near coord/.append style={rotate=90, anchor=west}
        ]
        \addplot table[x=Edge similarity measures,y=NDES]{\lookm};
        \addplot table[x=Edge similarity measures,y=Cosine]{\lookm};
        \addplot table[x=Edge similarity measures,y=Jaccard]{\lookm};
        \addplot table[x=Edge similarity measures,y=Adamic adar]{\lookm};
         \legend{NDES,Cosine,Jaccard,Adamic adar}
  ,  \end{axis}
\end{tikzpicture}
\caption{\footnotesize Comparative quality analysis of edge similarity measures on Riskmap dataset}
\label{figfive}
\end{figure}
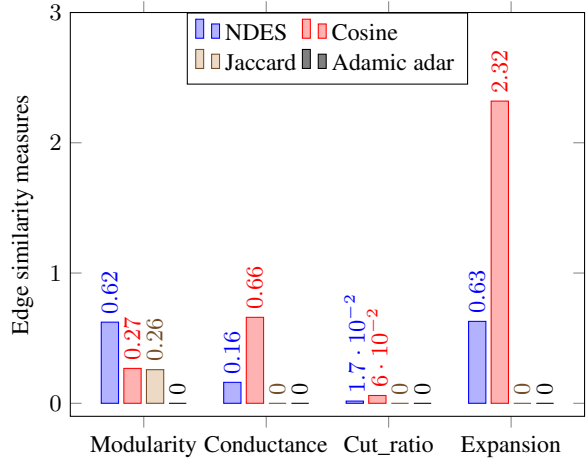

\pgfplotstableread[row sep=\\,col sep=&]{
Edge similarity measures & NDES & Cosine & Jaccard & Adamic adar\\
Modularity & 0.495 & 0.191 & 0.0 & 0.0 \\
Conductance & 0.251 & 0.742 & 0.0 & 0.0\\
Cut\_ratio & 0.027 & 0.081 & 0.0 & 0.0\\
Expansion & 1.29 & 4.76 & 0.0 & 0.0\\
}\lookm

\begin{figure}[!hb]
\centering
\begin{tikzpicture}
    \begin{axis}[
            ybar,
           bar width=.23cm,
            width=0.69\textwidth,
            height=.57\textwidth,
            enlarge x limits=0.2,
            legend style={at={(0.5,1)},
                anchor=north,legend columns=3,legend cell align=left},
            symbolic x coords={Modularity,Conductance,Cut\_ratio,Expansion},
            xtick=data,
             x tick label style={rotate=00,anchor=north},
            nodes near coords align={vertical},
             ymin=-0.10, ymax=5.9,
            ylabel={Edge similarity measures},
            nodes near coords,
            every node near coord/.append style={rotate=90, anchor=west}
        ]
        \addplot table[x=Edge similarity measures,y=NDES]{\lookm};
        \addplot table[x=Edge similarity measures,y=Cosine]{\lookm};
        \addplot table[x=Edge similarity measures,y=Jaccard]{\lookm};
        \addplot table[x=Edge similarity measures,y=Adamic adar]{\lookm};
         \legend{NDES,Cosine,Jaccard,Adamic adar}
  ,  \end{axis}
\end{tikzpicture}
\caption{\footnotesize Comparative quality analysis of edge similarity measures on Dolphin dataset}
\label{figsix}
\end{figure}
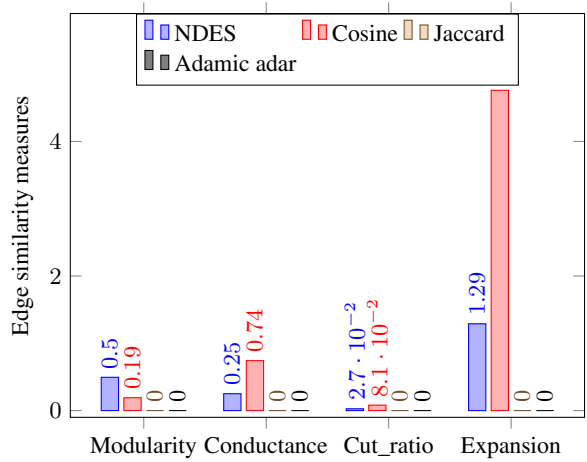

\subsection{Result Analysis}
\label{resultanalysis}

As a matter of primary importance, we have investigated efficiency of proposed measure in terms of accuracy with respect to several popular similarity measures. The comparative graphical results with respect to accuracy of NDES measure  on several real-world datasets are presented in Fig.~\ref{figone} and Fig.~\ref{figtwo}. From these figures, it is inferred that NDES gives comparatively good accuracy results on all the representative datasets. Though the accuracy have not been achieved fully, we have accomplished our goal better in comparison to other similarity metrics. Furthermore, we have also analyzed quality related aspects during this analysis. As can be seen in Fig.~\ref{figthree}, Fig.~\ref{figfour}, Fig.~\ref{figfive} and Fig.~\ref{figsix}, NDES gives the best modularity score out of all the selected similarity measures on all the representative datasets which indicates goodness of NDES in terms of internal connections. Next, comparative analysis with respect to cut\_ratio  and expansion shows that NDES gives comparatively good performance with respect to external connections. Additionally, as can be seen from these figures, the results given by Jaccard index and Adamic Adar are zero and hence, is negligible  and therefore, not considered for analysis. Similarly, the comparative analysis of NDES with respect to conductance shows the goodness of NDES with respect to combination of internal and external connections. Therefore, it can be concluded that neighborhood density and information concept adopted by NDES is effective and efficient in determining similarity between connected nodes.

\section{Conclusion}
\label{seven}


In this paper, we developed a neighborhood density based asymmetric similarity measure called NDES considering both dense and sparse networks. It utilizes neighborhood density information associated with the connected node pairs to compute edge similarity score. An application of our proposed similarity measure for community detection in OSNs have been shown. For this, we have substituted the similarity measure of our selected community detection technique with NDES. Extensive experiments on several small real-world datasets show that NDES achieves best accuracy and quality among Salton Index, Jaccard Coefficient and Adamic Adar. In addition to this, NDES has least time complexity in comparison to the baseline similarity measures. In future works, we plan to perform extensive experiments on large networks to evaluate the performance of NDES.


\begin{thebibliography}{6}
\bibitem{palla2005uncovering}
G. Palla, I. Der{\'e}nyi, I. Farkas, and T. Vicsek, “Uncovering the overlapping community structure of complex networks in nature and society,”nature, vol. 435, no. 7043, pp. 814–818, 2005.
\bibitem{lu2011link}
L. L{\"u} and T. Zhou, “Link prediction in complex networks: A survey,”Physica A: statistical mechanics and its applications, vol. 390, no. 6,pp. 1150–1170, 2011.

\bibitem{al2011survey}
M.  Al  Hasan  and  M.  J.  Zaki,  “A  survey  of  link  prediction  in  social networks,” in Social network data analytics.   Springer, 2011, pp. 243–275.

\bibitem{martinez2016survey}
V. Mart{\'\i}nez, F. Berzal, and J.-C. Cubero, “A survey of link prediction in complex networks,”ACM computing surveys (CSUR), vol. 49, no. 4,pp. 1–33, 2016.

\bibitem{dunlavy2011temporal}
 D.  M.  Dunlavy,  T.  G.  Kolda,  and  E.  Acar,  “Temporal  link  prediction using matrix and tensor factorizations,”ACM Transactions on Knowledge Discovery from Data (TKDD), vol. 5, no. 2, pp. 1–27, 2011.


\bibitem{liben2007link}
D.  Liben-Nowell  and  J.  Kleinberg,  “The  link-prediction  problem  for social  networks,”Journal  of  the  American  society  for  information science and technology, vol. 58, no. 7, pp. 1019–1031, 2007.





\bibitem{jaccard1901etude}
P.  Jaccard,  “ {\'E}tude  comparative  de  la  distribution  florale  dans  uneportion  des  alpes  et  des  jura,”Bull  Soc  Vaudoise  Sci  Nat,  vol.  37,pp. 547–579, 1901.

\bibitem{adamic2003friends}
 L. A. Adamic and E. Adar, “Friends and neighbors on the web,”Social networks, vol. 25, no. 3, pp. 211–230, 2003.
 
\bibitem{zhou2009predicting}
T.  Zhou,  L.  L {\"u},  and  Y.-C.  Zhang,  “Predicting  missing  links  via  local information,”The  European  Physical  Journal  B,  vol.  71,  no.  4,  pp.623–630, 2009.
 
\bibitem{sorensen1948method}
T. A. Sorensen, “A method of establishing groups of equal amplitude in  plant  sociology  based  on  similarity  of  species  content  and  itsapplication  to  analyses  of  the  vegetation  on  danish  commons,”Biol.Skar., vol. 5, pp. 1–34, 1948.



\bibitem{salton1983introduction}
 G.  Salton  and  M.  J.  McGill, Introduction  to  modern  information retrieval.    mcgraw-hill, 1983.

\bibitem{tan2014link}
F.  Tan,  Y.  Xia,  and  B.  Zhu,  “Link  prediction  in  complex  networks:  a mutual  information  perspective,”PloS  one,  vol.  9,  no.  9,  p.  e107056,2014.

\bibitem{barabasi2002evolution}
A.-L.  Barab{\^a}si,  H.  Jeong,  Z.  N{\'e}da,  E.  Ravasz,  A.  Schubert,  andT. Vicsek, “Evolution of the social network of scientific collaborations,”Physica A: Statistical mechanics and its applications, vol. 311, no. 3-4,pp. 590–614, 2002.

\bibitem{ghorbanzadeh2021hybrid}
H.  Ghorbanzadeh,  A.  Sheikhahmadi,  M.  Jalili,  and  S.  Sulaimany,  “Ahybrid  method  of  link  prediction  in  directed  graphs,”Expert  Systemswith Applications, vol. 165, p. 113896, 2021.



\bibitem{aghabozorgi2018new}
F. Aghabozorgi and M. R. Khayyambashi, “A new similarity measure for  link  prediction  based  on  local  structures  in  social  networks,”Physica  A:  Statistical  Mechanics  and  its  Applications,  vol.  501,  pp.12–23, 2018.


\bibitem{li2018similarity}
S. Li, J. Huang, Z. Zhang, J. Liu, T. Huang, and H. Chen, “Similarity-based future common neighbors model for link prediction in complex networks,”Scientific reports, vol. 8, no. 1, pp. 1–11, 2018.



\bibitem{jiang2013efficient}
Y. Jiang, C. Jia, and J. Yu, “An efficient community detection method based  on  rank  centrality,”Physica  A:  statistical  mechanics  and  its applications, vol. 392, no. 9, pp. 2182–2194, 2013.

\bibitem{li2015parameter}
Y. Li, C. Jia, and J. Yu, “A parameter-free community detection methodbased  on  centrality  and  dispersion  of  nodes  in  complex  networks,”Physica  A:  Statistical  Mechanics  and  its  Applications,  vol.  438,  pp.321–334, 2015.

\bibitem{wang2015novel}
T. Wang, H. Wang, and X. Wang, “A novel cosine distance for detecting communities  in  complex  networks,”Physica  A:  Statistical  Mechanics and its Applications, vol. 437, pp. 21–35, 2015.

\bibitem{eustace2015community}
 J. Eustace,  X.  Wang,  and  Y.  Cui,  “Community  detection  using  local neighborhood in complex networks,”Physica A: Statistical Mechanics and its Applications, vol. 436, pp. 665–677, 2015.

\bibitem{zhou2003distance}
H.   Zhou,   “Distance,   dissimilarity   index,   and   network   community structure,”Physical review e, vol. 67, no. 6, p. 061901, 2003.

\bibitem{clauset2005finding}
 A. Clauset, “Finding local community structure in networks,”Physical review E, vol. 72, no. 2, p. 026132, 2005.


\bibitem{bagrow2005local}
J. P. Bagrow and E. M. Bollt, “Local method for detecting communities,”Physical Review E, vol. 72, no. 4, p. 046108, 2005.

\bibitem{raghavan2007near}
U. N. Raghavan, R. Albert, and S. Kumara, “Near linear time algorithm to  detect  community  structures  in  large-scale  networks,”Physical review E, vol. 76, no. 3, p. 036106, 2007.

\bibitem{das2021community}
S.  Das  and  A.  Biswas,  “Community  detection  in  social  networks using local topology and information exchange,” in2021 International Conference on Intelligent Technologies (CONIT).   Hubli, India: IEEE,Jun. 25–27, 2021, pp. 1–7.


\bibitem{das2021deployment}
S.  Das  and  A.  Biswas, “Deployment of information diffusion for community detection in online social networks: A comprehensive review,”IEEE Transactions on Computational Social Systems, vol. 8, no. 5, pp. 1083–1107, 2021.

\bibitem{chakraborty2017}
T. Chakraborty, A. Dalmia, A. Mukherjee, and N. Ganguly, “Metrics
for community analysis: A survey,” ACM Computing Surveys (CSUR),
vol. 50, no. 4, pp. 1–37, Aug. 2017.



\bibitem{snap}
SNAP  Datasets:  Stanford  large  network  dataset  collection,”  http://snap.stanford.edu/data, Accessed September 27, 2021.

 








\end{thebibliography}
\end{document}